\documentclass[twoside,a4paper,11pt]{sca}
\usepackage{graphicx}
\usepackage{hyperref}
\usepackage{movie15}
\usepackage[belowskip=-15pt,aboveskip=0pt]{caption}

\begin{document}
\pagenumbering{arabic}
\pagestyle{myheadings}
\thispagestyle{empty}
{\flushright\includegraphics[width=\textwidth,bb=90 650 520 700]{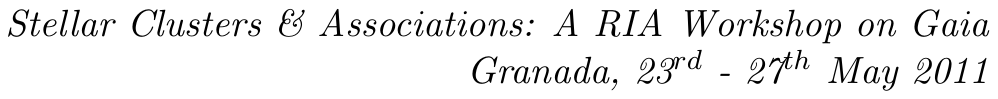}}
\vspace*{0.2cm}
\begin{flushleft}
{\bf {\LARGE
 
Feedback Regulated Star Formation: From Star Clusters to Galaxies
%
}\\
\vspace*{1cm}

Sami Dib$^{1}$
%
}\\
\vspace*{0.5cm}

$^{1}$
Astrophysics Group, Blackett Laboratory, Imperial College London, London, SW7 2AZ, United Kingdom; s.dib@imperial.ac.uk\\
%
%
%
\end{flushleft}
%
\markboth{
%
Feedback Regulated Star Formation
}{ 
%
Sami Dib
%
}
\thispagestyle{empty}
\vspace*{0.4cm}
\begin{minipage}[l]{0.09\textwidth}
\ 
\end{minipage}
\begin{minipage}[r]{0.9\textwidth}
\vspace{1cm}
\section*{Abstract}{\small
%
%

This paper summarises results from semi-analytical modelling of star formation in protocluster clumps of different metallicities. In this model, gravitationally bound cores form uniformly in the clump following a prescribed core formation efficiency per unit  time. After a contraction timescale which is equal to a few times their free-fall times, the cores collapse into stars and populate the IMF. Feedback from the newly formed OB stars is taken into account in the form of stellar winds. When the ratio of the effective wind energy of the winds to the gravitational energy of the system reaches unity, gas is removed from the clump and core and star formation are quenched. The power of the radiation driven winds has a strong dependence on metallicity and increases with increasing metallicity. Thus, winds from stars in the high metallicity models lead to a rapid evacuation of the gas from the protocluster clump and to a reduced star formation efficiency, $SFE_{exp}$, as compared to their low metallicity counterparts. By combining $SFE_{exp}$ with the timescales on which gas expulsion occurs, we derive the metallicity dependent  star formation rate per unit time in this model as a function of the gas surface density $\Sigma_{g}$. This is combined with the molecular gas fraction in order to derive the dependence of the surface density of star formation $\Sigma_{SFR}$ on $\Sigma_{g}$. This feedback regulated model of star formation reproduces very well the observed star formation laws extending from low gas surface densities up to the starburst regime. Furthermore, the results show a dependence of $\Sigma_{SFR}$ on metallicity over the entire range of gas surface densities, and can also explain part of the scatter in the observations. 

%
\normalsize}
\end{minipage}
%
%
%
%

\section{THE SFE IN STELLAR CLUSTERS AND IN GALAXIES}\label{sfe}

\begin{figure}[t]
\includegraphics[totalheight=0.5\textheight,width=\textwidth]{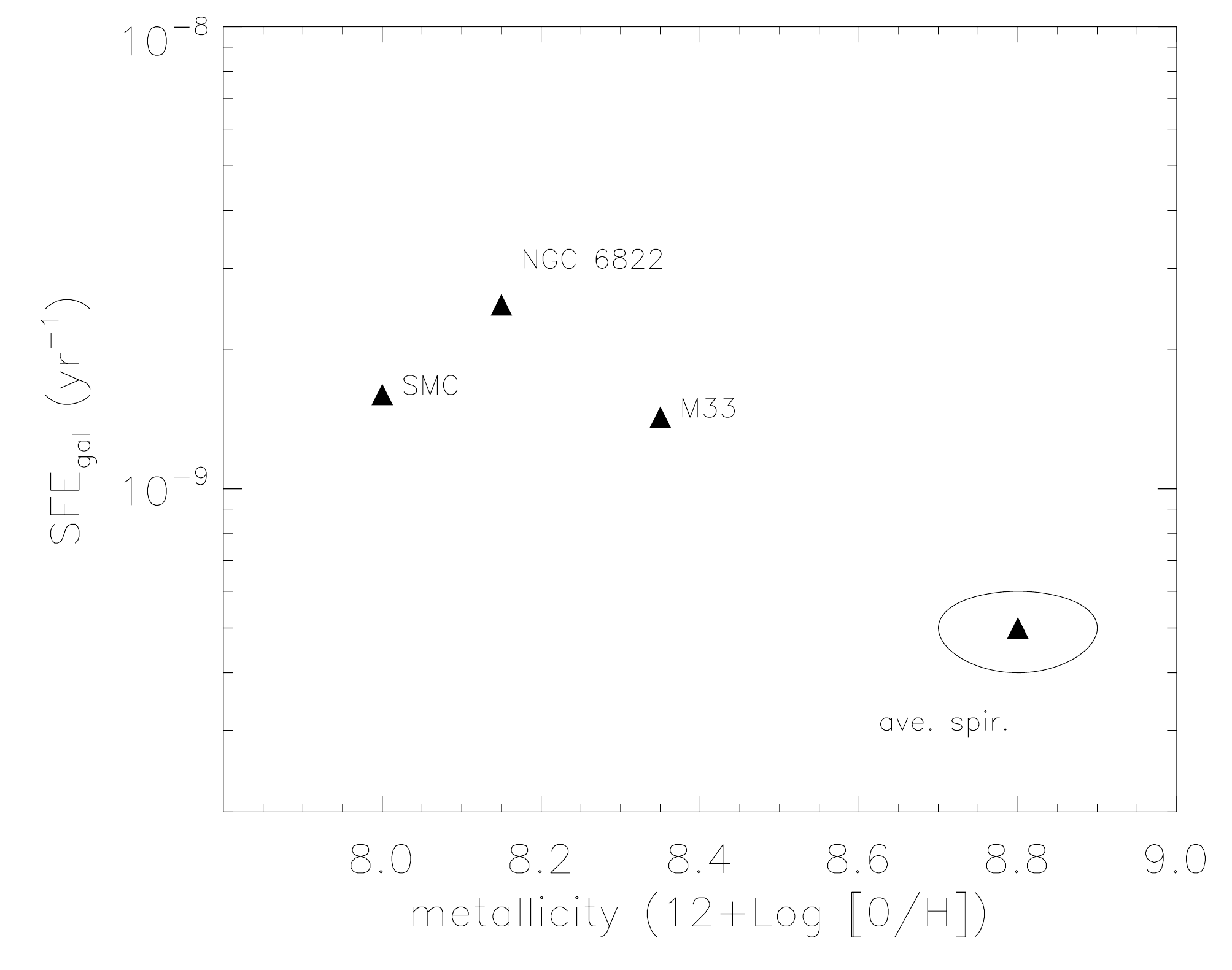} 
\caption{\label{fig1} {\small Global galactic star formation efficiency $SFE_{gal}$ as a function of the galactic global metallicity. The data for M33 and NGC 6822 are from Gratier et al. (2010a,b), for the SMC from Leroy et al. (2006), and the average spiral value from the sample of Murgia et al. (2002). Adapted from Dib et al. (2011)}
}
\end{figure}

The star formation efficiency (SFE) is one of the essential quantities that regulates the dynamical evolution and chemical enrichment of stellar clusters, the interstellar medium, and galaxies (e.g.,  Boissier et al. 2001; Krumholz \& McKee 2005; Dib et al. 2006; Dib et al. 2009; Dib et al. 2011). The SFE is commonly defined as being the fraction of gas which is converted into stars in a system of a given mass, be it a protocluster molecular clump, an entire giant molecular cloud (GMC), or a galaxy. In an isolated clump, the time evolving SFE is usually defined as being:

\begin{equation}
SFE (t) = \frac{M_{cluster}(t)}{M_{clump}},
\label{eq1}
\end{equation}

where $M_{cluster} (t)$ is the cluster mass at time $t$ and $M_{clump}$ is the initial clump mass. In observed protocluster clumps with embedded or semi-embedded clusters, the SFE is usually approximated by: 

\begin{equation} 
SFE_{obs}= \frac{M_{cluster}}{M_{cluster}+M_{gas}},
\label{eq2}
\end{equation}

\begin{figure}[tt]
\includegraphics[totalheight=0.55\textheight,width=\textwidth]{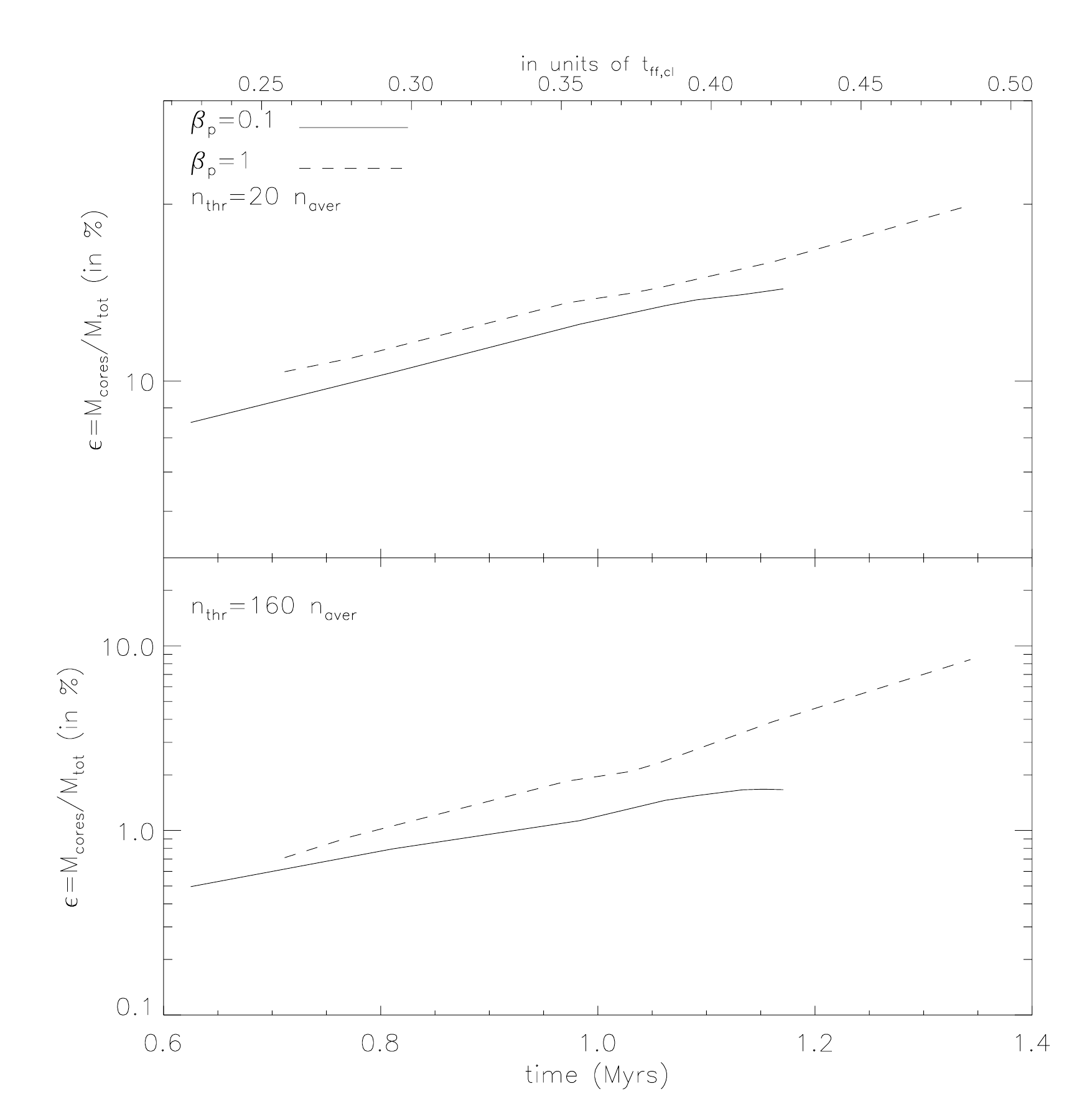} 
\caption{\label{fig2} \small{Time evolution of the fraction of cloud mass of the cloud that is in dense cores in two 3D numerical simulations of magnetized, turbulent, and self gravitating isothermal molecular cloud (Dib et al. 2010a). One simulation has a mass-to magnetic flux ratio of $\mu_{B}=2.8$ (mildly supercritical) and the other has $\mu_{B}=8.8$ (strongly supercritical). The upper corresponds to cores detected using a density threshold of $n_{thr}=20~n_{aver}$ and the lower panel corresponds to cores detected at the density threshold of $160~n_{aver}$, where $n_{aver}=500$ cm$^{-3}$ is the average number density in the cloud. The cores identified at the threshold density of $160–_{aver}$ corresponds to a population of gravitationally bound cores. Adapted from Dib et al. (2010a)
}}
\end{figure}

\noindent where $M_{gas}$ is the mass of the star forming gas. Measured SFEs of nearby embedded clusters, using Eq.~\ref{eq2} yield values that fall in the range 0.1-0.5 (Wilking \& Lada 1983; Rengarajan 1984; Wolf et al. 1990; Pandey et al. 1990; Lada et al. 1991a,1991b; Warin et al. 1996; Olmi \& Testi 2002). These SFE values are larger than those obtained for entire GMCs which are observed to fall in the range of a few percent (e.g., Duerr et al. 1982, Fukui \& Mizuno 1991; Evans et al. 2009). On galactic scales, the star formation efficiency is usually defined as being the inverse of the molecular gas consumption time and is given by:

\begin{equation} 
SFE_{gal}= \frac{SFR_{gal}} {M_{H_{2}}},
\label{eq3}
\end{equation}

\noindent where $SFR_{gal}$ is the galactic star formation rate (in M$_{\odot}$ yr$^{-1}$), and $M_{H_{2}}$ the mass of the molecular hydrogen gas. In large spiral galaxies $SFE_{gal} \sim 0.5\times 10^{-9}$ yr$^{-1}$ (e.g., Kennicutt 1998, Murgia et al. 2002; Leroy et al. 2008). Fig.~\ref{fig1} displays the global $SFE_{gal}$ for a few selected nearby galaxies as a function of their global metallicity. The trend in Fig.~\ref{fig1} is strongly suggestive of an $SFE_{gal}$ which increases with decreasing metallicity. Several physical processes regulate the core formation efficiency (CFE) within a star forming molecular clump/cloud, and as a consequence, the SFE. Supersonic turbulence can provide support against the global collapse of the clump/cloud, or at least delay it (e.g., V\'{a}zquez-Semadeni \& Passot 1999). However, on scales smaller than the energy injection scale, but larger than the sonic scales in the cloud, supersonic turbulence produces local compressions (i.e., cores) (e.g., Padoan 1995; V\'{a}zquez-Semadeni et al. 2003; Dib et al. 2010a) of which a fraction can be 'captured' by gravity and proceed to collapse into stars (e.g., V\'{a}zquez-Semadeni et al. 2005a; Dib et al. 2007a; Dib \& Kim 2007; Dib et al. 2008a; Dib et al. 2010a). Krumholz \& McKee (2005) formulated an analytical theory in which the core formation efficiency per free-fall time, $CFE_{ff}$, is shown to decrease with an increasing sonic Mach number and an increasing virial parameter and is given by $CFE_{ff}\approx 0.15 \alpha_{vir}^{-0.68} {\cal{M}}^{-0.32}$.

\begin{figure}[t]
\includegraphics[totalheight=0.45\textheight,width=\textwidth]{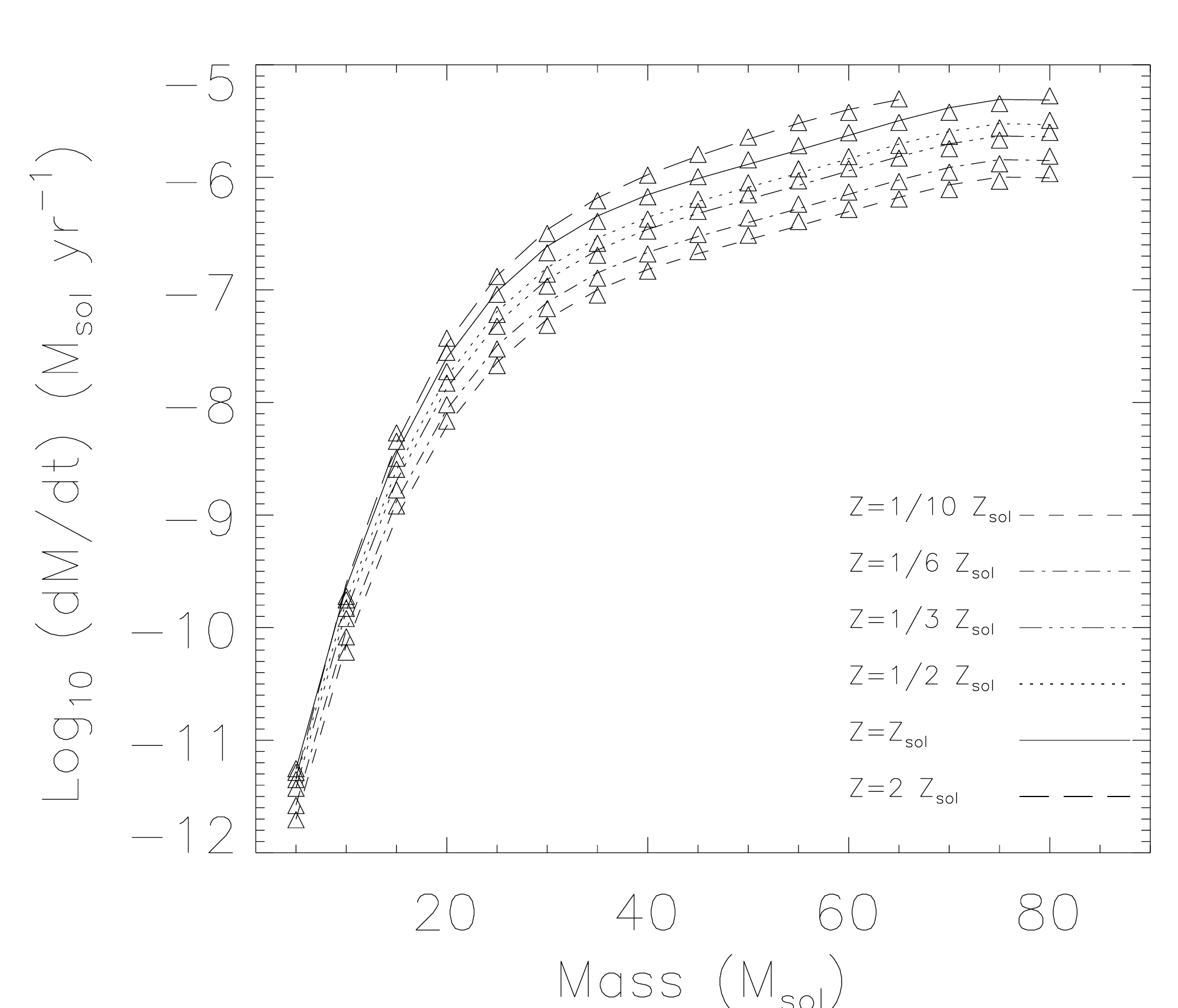} 
\caption{\label{fig3} {\small Stellar mass loss rates for stars in the mass range 5-80 M$_{\odot}$ on the main sequence, and for various metallicities. The stellar mass loss rates have been calculated using the stellar characteristics (effective temperature, stellar luminosity and radius) computed using the stellar evolution code CESAM coupled to the stellar atmosphere model of Vink et al. (2001). Over-plotted to the data are fourth order polynomial fits. The parameters of the fit functions can be found in Dib et al. (2011). Adapted from Dib et al. (2011)}
}
\end{figure}

\begin{figure}[t]
\includegraphics[totalheight=0.5\textheight,width=\textwidth]{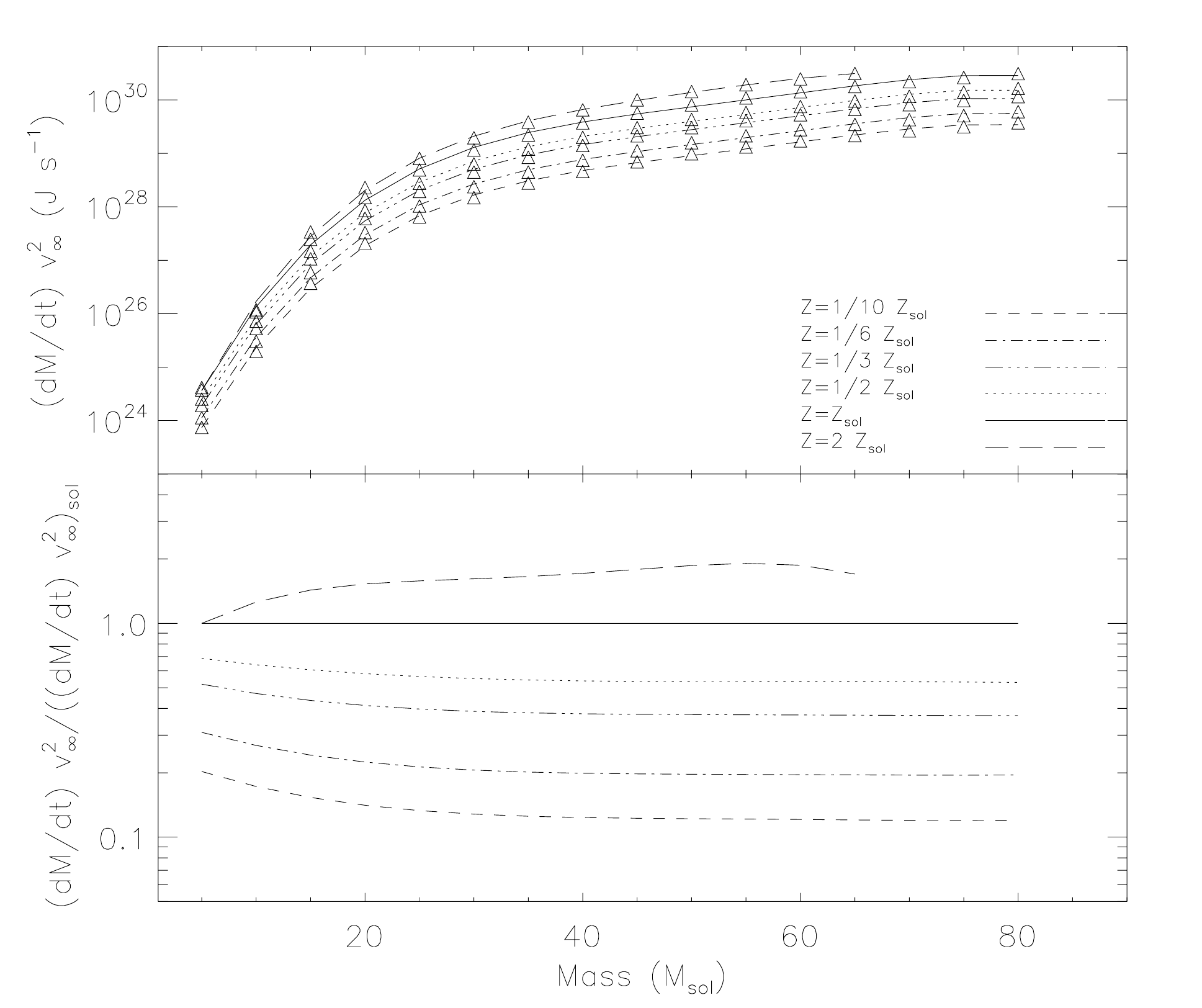} 
\caption{\label{fig4} {\small The power of the stellar winds, or wind luminosities, for stars in the mass range 5-80 M$_{\odot}$ on the main sequence, and for various metallicities. The stellar mass loss rates have been calculated using the stellar characteristics (effective temperature, stellar luminosity and radius) computed using the stellar evolution code CESAM coupled to the stellar atmosphere model of Vink et al. (2001). The values of $v_{\infty}$ have been calculated using the derivation by Leitherer et al. (1992). Over-plotted to the data are fourth order polynomials. The parameters of the fit functions can be found in Dib et al. (2011). Adapted from Dib et al. (2011)}
}
\end{figure}

Magnetic fields play an important role in determining the fraction of gravitationally bound gas in star forming clouds/clumps. Results from numerical simulations show that stronger magnetic fields (in terms of magnetic criticality) lower the rate of dense core formation in a star forming molecular clump/cloud (e.g., V\'{a}zquez-Semadeni et al. 2005b; Price \& Bate 2008; Dib et al. 2008a; Li et al. 2010; Dib et al. 2010a). Dib et al. (2010a) showed that the CFE per unit of the free-fall time of the cloud, $CFE_{ff}$, are of the order of $\sim 6~\%$ and $\sim 33~\%$ for clouds with mass to-magnetic flux ratios of $\mu=2.2$ and $8.8$, respectively (with $\mu$ being normalised by the critical mass-to-flux ratio for collapse, see Fig.~\ref{fig2}). The role of stellar feedback in setting the final value of the SFE has been investigated by several authors. The role of protostellar outflows has been studied theoretically by Adams \& Fatuzzo (1996), Matzner \& McKee (2000) and numerically by Nakamura \& Li (2007) and Li et al. (2010). Supernova explosions are an efficient way of removing gas from the protocluster region (e.g., Parmentier et al. 2008; Baumgardt et al. 2008). However, they occur after a few million years from the time massive stars have formed. Another form of stellar feedback is associated with O and B stars, in their main sequence phase, and eventually beyond. OB stars emit UV radiation which ionises the surrounding gas and heats it to temperatures of $\sim 7000-10^{4}$ K. This warm and ionised bubble provides the environment in which particles accelerated from the stellar surface by interaction with some of the stellar radiation propagate outwards. In the following sections, we present a model which describes the co-evolution of the mass function of gravitationally bound cores and the IMF in a protocluster region. In the model, dense cores form in the protocluster clump uniformly in time following a specified core formation efficiency per unit free-fall time of the clump, $CFE_{ff}$. The dense cores have lifetimes of a few times their free-fall times, after which they collapse to form stars and populate the IMF. Stellar winds from the newly formed massive stars ($M_{\star} > 5~$M$_{\odot}$) inject energy into the clump and when the ratio of the effective wind energy to the gravitational energy of the clump reaches unity, gas is removed from the clump and core and star formation are quenched. We discuss the dependence of the final star formation efficiency on the metallicity. Finally, we present a derivation of the star formation laws in galaxies in the context of this metallicity dependent, feedback regulated model of star formation. 

\section{The Model}

\subsection{Protocluster Clumps}\label{clusters}

Several studies have established that star clusters form in dense ($> 10^{3}$ cm$^{-3}$) clumps embedded in a lower density parental molecular cloud (e.g., Lada \& Lada 2003; Klein et al. 2005; Rathborne et al. 2006; Kauffmann \& Pillai 2010; Csengeri et al. 2011; Hernandez et al. 2011).  Saito et al. (2007) studied, using  the C$^{18}$O molecular emission line, a large sample of cluster forming clumps whose masses and radii vary between [15-1500] M$_{\odot}$ and [0.14-0.61] pc, respectively. The mass-size and velocity dispersion-size relations of clumps in the sample of Saito et al. (2007) were fitted by Dib et al. (2010b), obtaining: 

\begin{equation}
M_{clump}({\rm M_{\odot}})=10^{3.62 \pm 0.14} R_{c}^{2.54 \pm 0.25} ({\rm pc}), 
\label{eq4}
\end{equation}

\noindent and 

\begin{equation}
v_{c}({\rm km~s^{-1}})=10^{0.45 \pm 0.08} R_{c}^{0.44 \pm 0.14}({\rm pc}),
\label{eq5}
\end{equation}

\noindent where $M_{clump}$ is the mass of the clump, $R_{c}$ its radius, and $v_{c}$ the scale dependent gas velocity dispersion. In this work, we adopt a protocluster clump model that follows an $r^{-2}$ density profile:

\begin{equation} 
\rho_{c}(r)= \frac{\rho_{c0}} {1+(r/R_{c0})^{2}},
\label{eq6}
\end{equation}

\noindent where $R_{c0}$  is the clump's core radius, $\rho_{c0}$ is the density at the centre. For a given value of $M_{clump}$, the central density $\rho_{c0}$ is given by:

\begin{equation} 
\rho_{c0}= \frac{M_{clump}}{4 \pi R_{c0}^{3} [(R_{c}/R_{c0})-\arctan(R_{c}/R_{c0})]}.
\label{eq7}
\end{equation}

The temperatures of the cluster forming clumps are observed to vary between 15 and 70 K (e.g., Saito et al. 2007; Rathborne et al. 2010). In order to further constrain the models and minimise the number of parameters, we relate the sizes of the protocluster clumps to their masses using the mass-size relation of Saito et al.  In the absence of detailed information about the velocity dispersion inside the cores in the Saito et al. (2007) study, we assume that the clump-clump velocity dispersion they derived (i.e., Eq~\ref{eq6}) is also valid on the scale of the clumps themselves and of their substructure. 

\subsection{Prestellar Cores Model}\label{core_model}

Whitworth \& Ward-Thompson (2001) applied a family of Plummer sphere-like models to the contracting prestellar dense core L1554, which is representative of the population of gravitationally bound cores in clumps that are considered in this work. They found a good agreement with the observations of L1554 if the density profile of the core has the following form:

\begin{equation} 
\rho_{p}(r_{p})= \frac{\rho_{p0}}{[1+(r_{p}/R_{p0})^{2}]^{2}},
\label{eq8}
\end{equation}

\noindent where $\rho_{p0}$ and $R_{p0}$ are the central density and core radius of the core, respectively. Note that the radius of the core, $R_{p}$, depends both on its mass and on its position within the clump. The dependence of $R_{p}$ on $r$ requires that the density at the edges of the core equals the ambient clump density, i.e., $\rho_{p}(R_{p})=\rho_{c}(r)$. This would result in smaller radii for cores of a given mass when they are located in their inner parts of the clump. The density contrast between the centre of the core and its edge is given by: 

\begin{equation} 
{\cal C}(r) \equiv \frac {\rho_{p0}}{\rho_{c} (r)}=\frac {\rho_{p0}} {\rho_{c0}} \left[1+ \left(\frac{r}{R_{c0}}\right)^{2} \right].         
\label{eq9}
\end{equation}

Depending on its position $r$ in the clump, the radius of the core of mass $M$, $R_{p}$, can be calculated as being $R_{p} (r,M)=a(r)~R_{p0} (r,M)$, where: 

\begin{equation} 
R_{p0}(r,M)= \left(\frac{M}{2 \pi \rho_{p0}} \right)^{1/3} \left(\arctan[a(r)]-\frac{a(r)}{1+a(r)^{2}} \right)^{-1/3},
\label{eq10}
\end{equation} 

\begin{figure}[t]
\includegraphics[totalheight=0.60\textheight,width=\textwidth]{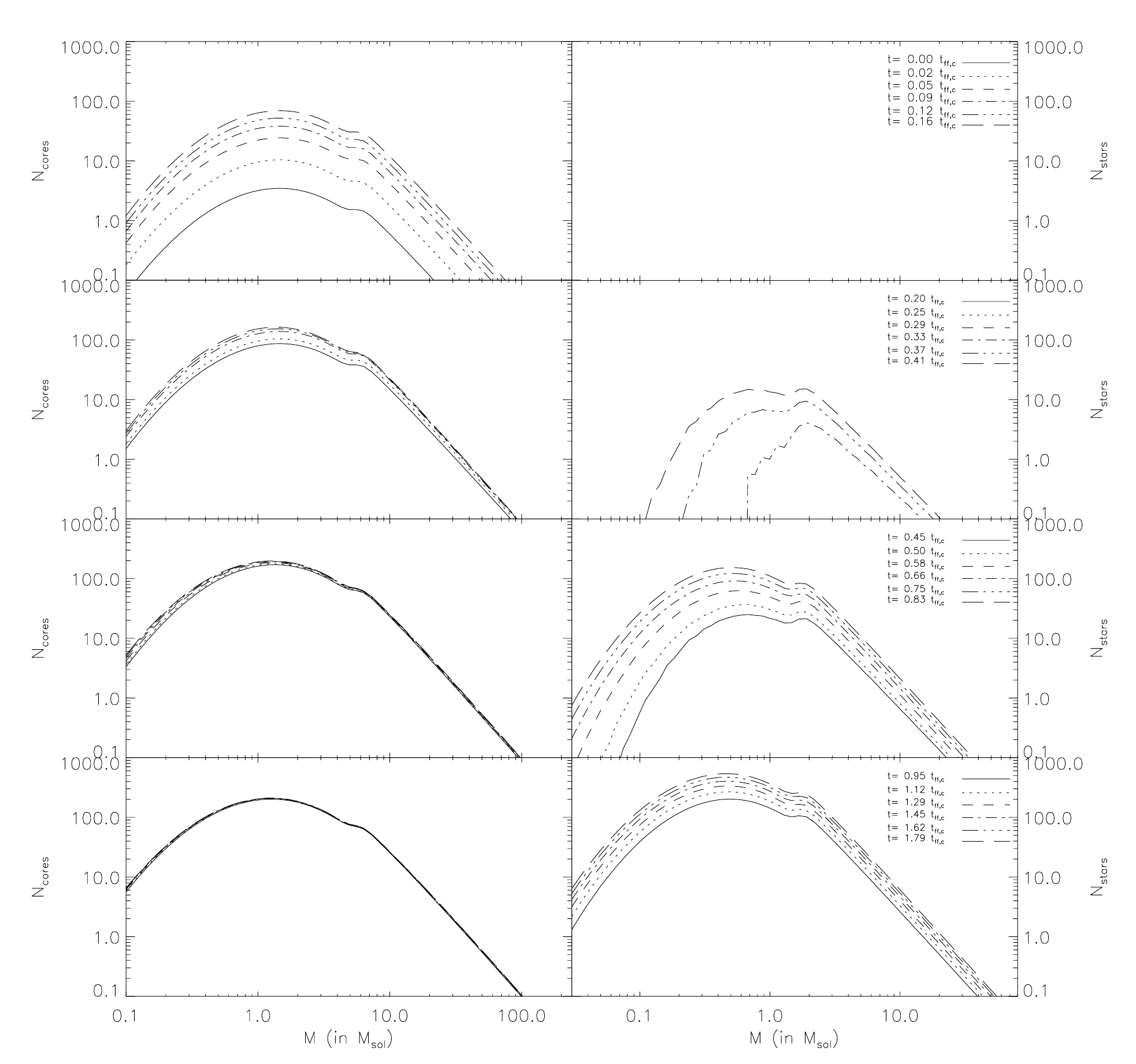} 
\caption{\label{fig5} {\small Time evolution of the pre-stellar core mass function (left), and stellar mass function (right) in the protocluster clump with the fiducial model parameters. The last  time-step shown is $1.79~t_{ff,c}$ which corresponds to the epoch at which gas is expelled from the protocluster clump. Adapted from Dib et al. (2011)}
}
\end{figure}

 \begin{figure}[t]
\includegraphics[totalheight=0.45\textheight,width=\textwidth]{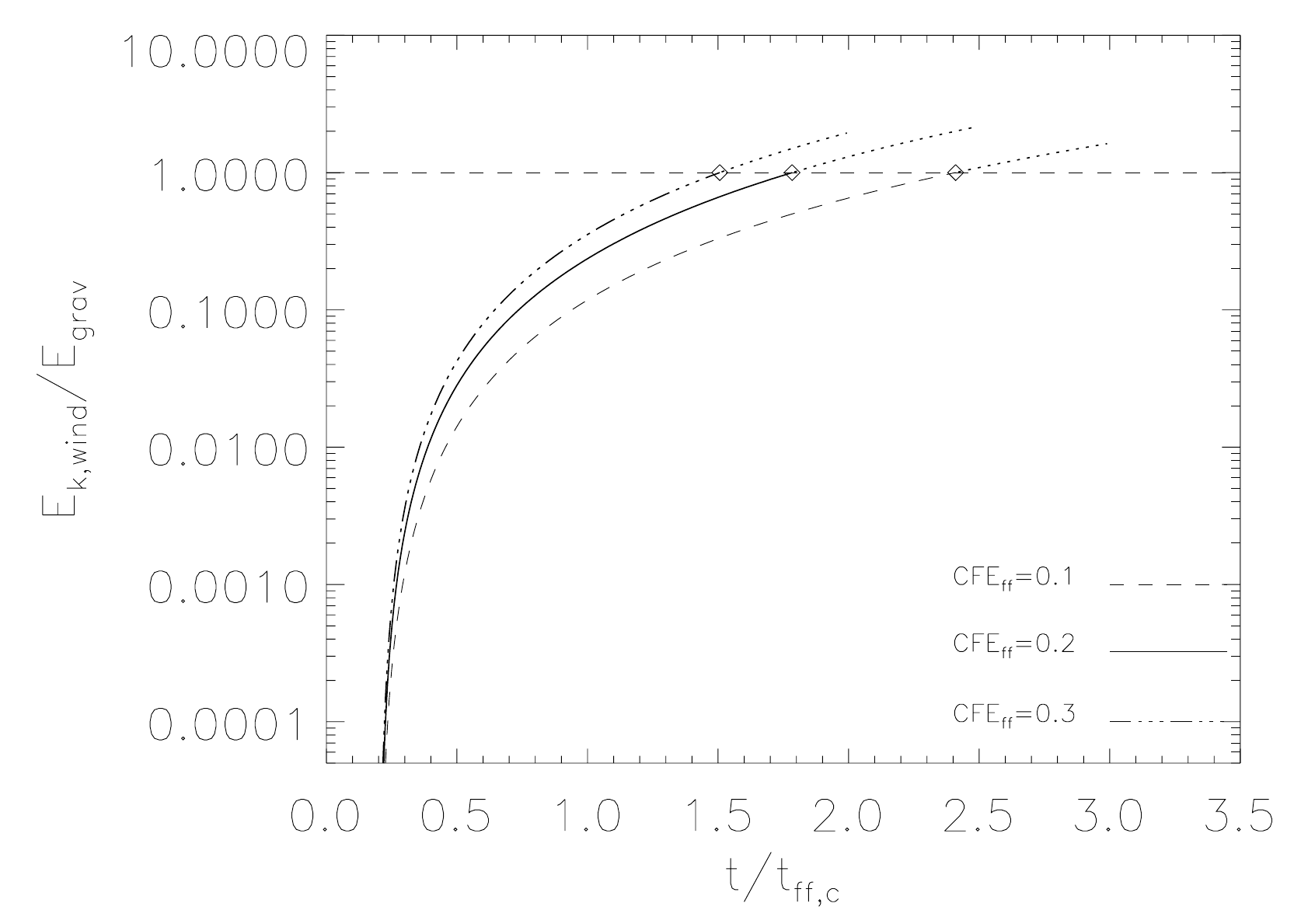} 
\caption{\label{fig6} {\small Time evolution of the ratio of the effective kinetic energy generated by stellar winds and the gravitational energy of the protocluster clump. Time is shown in units of the protocluster clump free fall timescale $t_{ff,c}$. The horizontal dashed corresponds to $E_{k,wind}/E_{grav} =1$ with  $\kappa =0.1$. The full line correspond to the fiducial model with the $CFE_{ff}=0.2$ and the dashed and triple dot-dashed to cases with $CFE_{ff}=0.1$ and $CFE_{ff}=0.3$, respectively.  Diamonds correspond to the epochs at which the gas is evacuated from the cluster in the three models and the processes of core and star formation are terminated. Adapted from Dib et al. (2011).}
} 
\end{figure}

\noindent and with $a(r) \equiv ({\cal C}(r)^{1/2}-1)^{1/2}$. With our set of parameters, the quantity ${\cal C}^{1/2}-1$ is always guaranteed to be positive. The value $R_p(r,M)$ can be considered as being the radius of the core at the moment of its formation. The radius of the core will decrease as time advances due to gravitational contraction. We assume that the cores contract on a timescale, $t_{cont,p}$ which we take to be a few times their free fall timescale $t_{ff,core}$, and which is parametrized by $t_{cont,p}(r,M)= \nu ~ t_{ff,core}(r,M)= \nu \left( 3 \pi/32~G \bar{\rho_{p}} (r,M) \right)^{1/2}$, where $G$ is the gravitational constant, $\nu$ is a constant $\ge 1$ and $\bar{\rho_{p}}$ is the radially averaged density of the core of mass $M$, located at position $r$ in the clump. The time evolution of the radius of a core of mass $M$, located at position $r$ in the cloud is given by a simple contraction law $R_{p}(r,M,t)=R_p(r,M,0)~e^{-(t/t_{cont,p})}$.  Both observational (Jessop \& Ward-Thompson 2000; Ward-Thompson et al. 2007) and numerical (V\'{a}zquez-Semadeni et al. 2005a; Galv\'{a}n-Madrid et al. 2007; Dib et al. 2008c; Gong \& Ostriker 2011) estimates of gravitationally bound cores lifetimes tend to show that they are of the order of a few times their free-fall time. One important issue is the choice of the cores central density, $\rho_{p0}$. In this work, we first assume that the minimum density contrast that exists between the centre of the core and its edge is of the order of the critical Bonnor-Ebert value and that is $> 15$. Secondly, we assume that the density contrast between the centre and the edge of the cores depends on their masses following a relation of the type $\rho_{p0}  \propto M^{\mu}$. Thus, the density contrast between the centre and the edge for a core with the minimum mass we are considering, $M_{min}$ (typically $M_{min}=0.1$ M$_{\odot}$) is 15, whereas for a more massive core of mass $M$, the density contrast will be equal to $15\times (M/M_{min})^{\mu}$.  Observations show that  $\mu$ varies in the range $[0-0.6]$ (e.g., Johnstone \& Bally 2006).
 
\subsection{The Initial Prestellar Cores Distributions} \label{ini_cond}

We assume that dense cores form in the clump as a result of its gravo-turbulent fragmentation.The mass distributions of the core formed at every epoch is described using the formulation of  Padoan \& Nordlund (2002). Thus, the local mass distributions of cores, $N(r,M)$, are given by: 

\begin{eqnarray}
N (r,M)~d\log~M =f_{0}(r)~M^{-3/(4-\beta)} \nonumber \\
            \times \left[\int^{M}_{0} P(M_{J}) dM_{J}\right]d\log~M,
\label{eq11}
\end{eqnarray}   

\noindent where $\beta$ is the exponent of the kinetic energy power spectrum, $E_{k} \propto k^{-\beta}$, and is related to the exponent $\alpha$ of the size-velocity dispersion relation in the clump with $\beta=2 \alpha+1$.  The local normalisation coefficient $f_{0}(r)$ is obtained by requiring that $\int^{M_{max}}_{M_{min}} N (r,M)~dM=1$ in a shell of width $dr$, located at distance $r$ from the clump's centre. $P(M_{J})$ is the local distribution of Jeans masses given by:

\begin{equation}
P(M_{J})~dM_{J}=\frac{2~M_{J0}^{2}}{\sqrt{2 \pi \sigma^{2}_{d}}} M^{-3}_{J} \exp \left[-\frac{1}{2} \left(\frac{\ln~M_{J}-A}{\sigma_{d}} \right)^{2} \right] dM_{J},
\label{eq12}
\end{equation}   

\noindent where $M_{J0}$ is the Jeans mass at the mean local density, and $\sigma_{d}$ is the standard deviation of the density distribution which is a function of the local thermal rms Mach number. Therefore, the local distribution of cores generated in the clump, at an epoch $\tau$, $N(r,M,\tau)$, is obtained by multiplying the local normalised function $N(r,M)$ by the local rate of fragmentation such that:

\begin{equation}
{N} (r,M,\tau) dt=\frac{CFE_{ff} (r) \rho_{c}(r)} {<M>(r)~t_{cont,p} (r,M)} \frac{dt}{t_{ff,c}} N(r,M),
\label{eq13}
\end{equation} 

\noindent where $dt$ is the time interval between two consecutive epochs, $<M>$ is the average core mass in the local distribution and is calculated by $<M>=\int_{M_{min}}^{M_{max}} M~N (r,M,0)~dM$, and $CFE_{ff}$ is a parameter smaller than unity which describes the local mass fraction of gas that is transformed into cores per free fall time of the protocluster clump, $t_{ff,c}$. In the present study, we assume that $CFE_{ff}$ is independent of $r$. 
 
 \section{Feedback model}\label{feedback}
 
\begin{figure}[ht]
\includegraphics[totalheight=0.45\textheight,width=\textwidth]{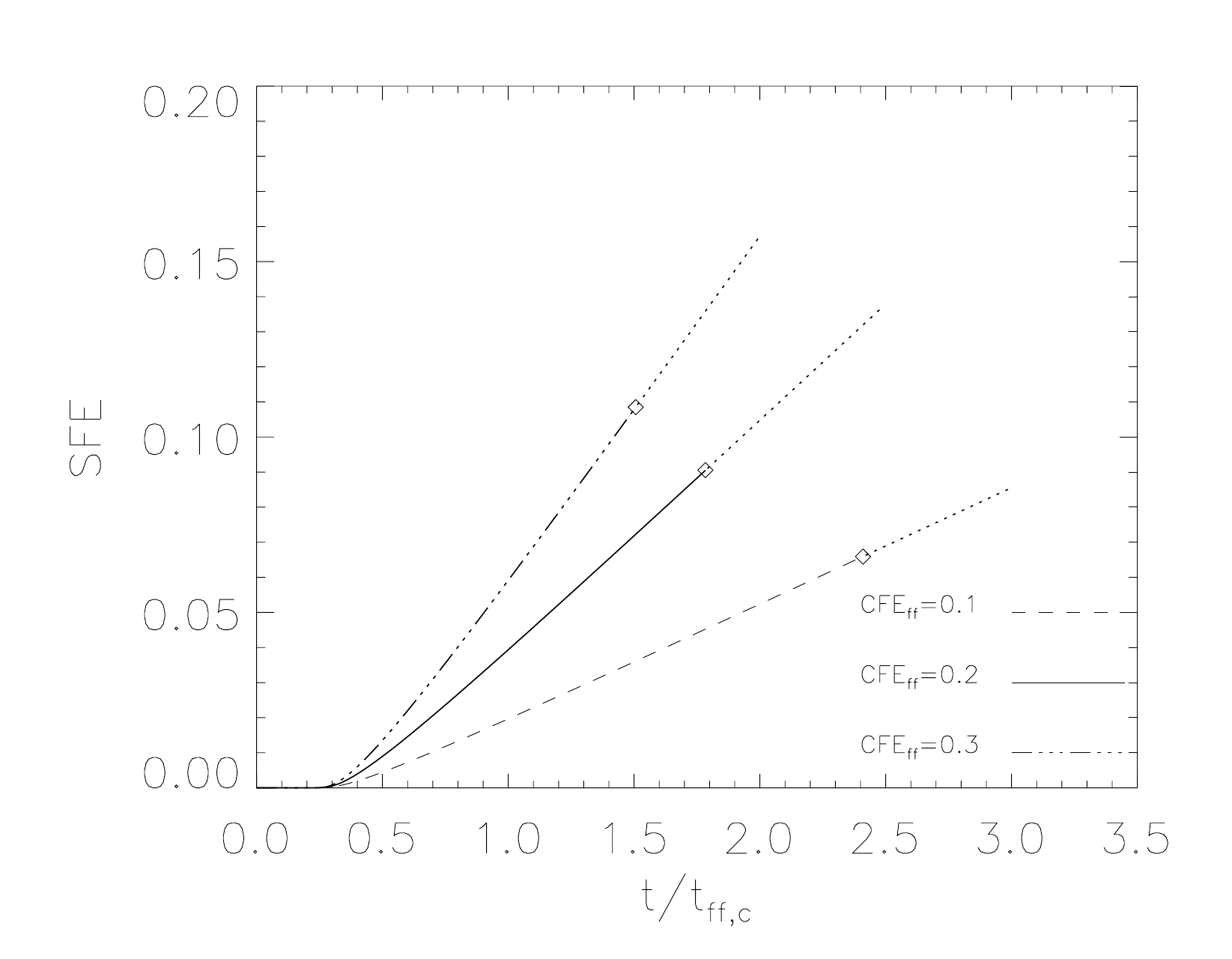} 
\caption{\label{fig7} {\small Time evolution of the SFE in the protocluster clump. The full line corresponds to the fiducial model with the $CFE_{ff}=0.2$ and the dashed and triple dot-dashed to cases with $CFE_{ff}=0.1$ and $CFE_{ff}=0.3$, respectively. Time is shown in units of the protocluster clump free fall timescale $t_{ff,c}$. Diamonds correspond to the epochs at which the gas is evacuated from the cluster and indicates the final value of the SFE, $SFE_{f}$, in the three models. Adapted from Dib et al. (2011).}
}
\end{figure}

\begin{figure}[ht]
\includegraphics[totalheight=0.45\textheight,width=\textwidth]{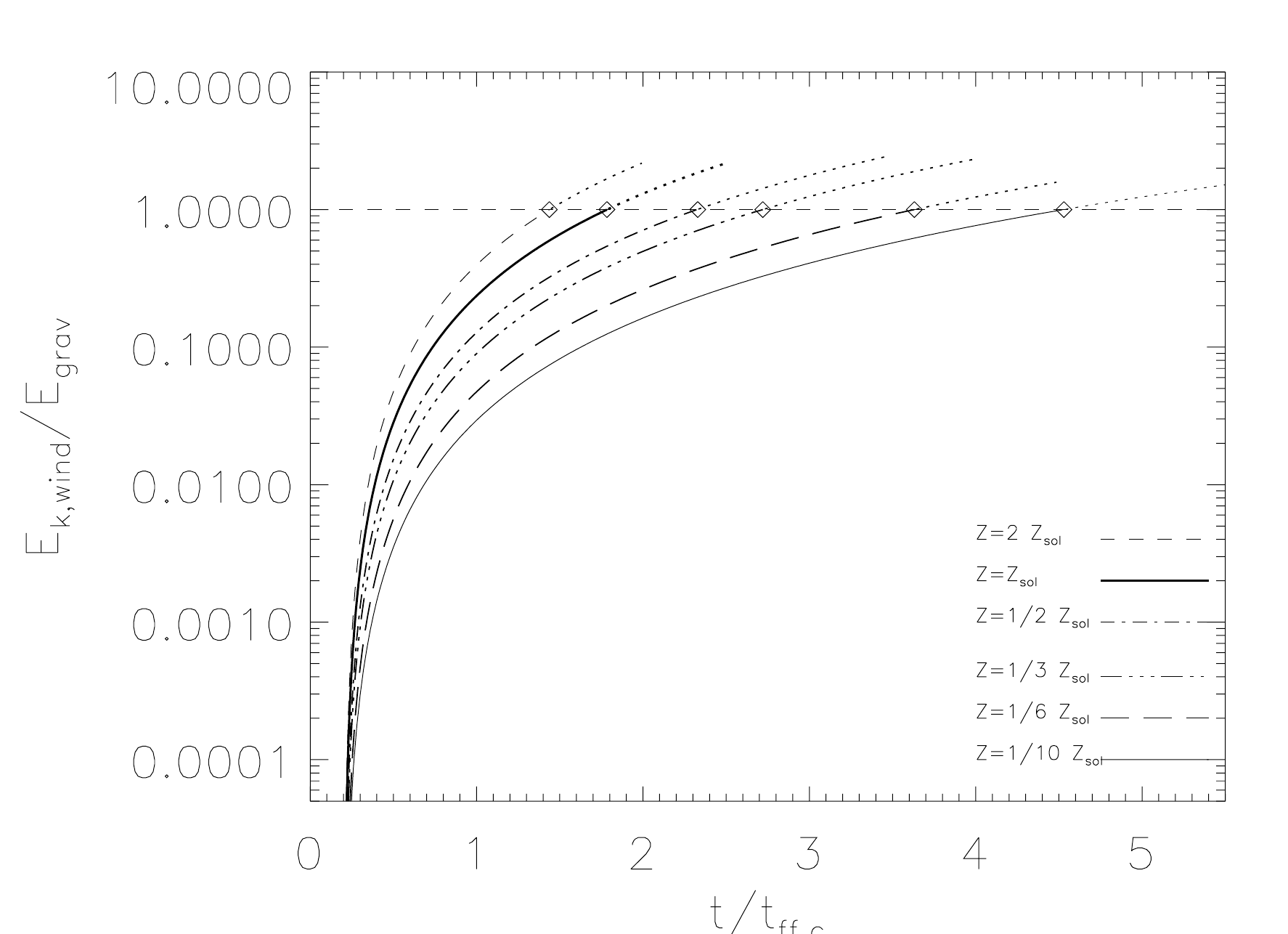} 
\caption{\label{fig8} {\small Time evolution of the ratio of the kinetic energy generated by stellar winds and the gravitational energy of the protocluster clump. Time is shown in units of the protocluster clump free fall timescale $t_{ff,c}$. The horizontal dashed corresponds to $E_{k,wind}/E_{grav} =1$ with $\kappa=0.1$. Diamonds correspond to the epochs at which the gas is evacuated from the cluster in the three models. The mass of the clump in these models is $10^{5}$ M$_{\odot}$. Adapted from Dib et al. (2011).}
}
\end{figure}
 
 In this model, the formation of cores in the protocluster clump, and consequently star formation, are terminated whenever the fraction of the wind energy stored into motions that oppose gravity exceeds the gravitational energy of the clump.  Thus, at any epoch $t < t_{exp}$ ($t_{exp}$ is the epoch at which gas is expelled from the clump), gas is removed from the clump only to be turned into stars. We take into account the feedback generated by the stellar winds of massive stars ($M_{\star} \ge 5$ M$_{\odot}$). In order to calculate reliable estimates of the feedback generated by metallicity dependent stellar winds, we proceed in two steps. In the first step, we use a modified version of the stellar evolution code CESAM (see appendix 1 in Piau et al. 2011) to calculate a grid of main sequence stellar models for stars in the mass range [5-80] M$_\odot$ (with steps of 5 M$_{\odot}$) at various metallicities $Z/Z_{\odot}=[1/10, 1/6, 1/3, 1/2, 1, 2]$ ($Z_{\odot}=0.0138$). The evolution of massive stars is followed using the CESAM code for $\sim 1$ Myr , on the main sequence. The characteristic stellar properties, which are the effective temperature $T_{eff}$, the luminosity $L_{\star}$, and the stellar radius $R_{\star}$ are then used in the stellar atmosphere model of  Vink et al. (2001) in order to calculate the stellar mass loss rate $\dot{M}_{\star}$. Vink et al. (2001) did not derive the values of the terminal velocities of the winds ($v_{\infty}$), therefore, we use instead the derivations of $v_{\infty}$ obtained by Leitherer et al. (1992). Fig.~\ref{fig3} displays the mass loss rates calculated for OB stars at the various metallicities. The power of the stellar winds is given by $\dot{M}_{\star} v_{\infty}^{2}$. This quantity is displayed in Fig.~\ref{fig4} for the models with different metallicities. Both ${\rm log} (\dot{M}_{\star})$ and $\dot{M}_{\star} v_{\infty}^{2}$ are fitted with fourth order polynomials (overplotted to the data) and whose coefficients are provided in Dib et al. (2011). The $\dot{M}_{\star} v_{\infty}^{2}-M_{\star}$ relations displayed in Fig.~\ref{fig4} allow for the calculation of the total wind energy deposited by stellar winds. The total energy from the winds is given by:

\begin{equation}
E_{wind} = \int_{t'=0}^{t'=t} \int_{M_{\star}=5~\rm{M_{\odot}}}^{M_{\star}=120~\rm{M_{\odot}}} \left( \frac{N(M_{\star}) \dot{M_{\star}} (M_{\star}) v_{\infty}^{2}}{2} dM_{\star}\right) dt'. 
\label{eq14}
\end{equation}

We assume that only a fraction of $E_{wind}$ will be transformed into systemic motions that will oppose gravity and participate in the evacuation of the bulk of the gas from the proto-cluster clump. The rest of the energy is assumed to be dissipated in wind-wind collisions or escape the wind bubble. The effective kinetic wind energy is thus given by: 

\begin{equation}
 E_{k,wind}=\kappa~E_{wind},
\label{eq15}
\end{equation}
 
\noindent where $\kappa$ is a quantity $\leq 1$. It is currently difficult to estimate $\kappa$ as its exact value will vary from system to system depending on the number of massive stars, their locations, and their wind interactions. As a conservative guess for the fiducial model, we take $\kappa=0.1$. $E_{k,wind}$ is compared at every timestep to the absolute value of the gravitational energy, $E_{grav}$, which is calculated as being:

\begin{equation}
E_{grav} =  -\frac{16}{3} \pi^{2} G \int_{0}^{R_{c}} \rho_{c}(r)^{2} r^{4}  dr,
\label{eq16}
\end{equation}

\noindent where $\rho_{c}$ is given by Eq.~\ref{eq6}.

\subsection{The co-evolution of the core mass function and of the IMF}\label{co_evol_fiducial}

Whenever a population of cores of mass $M$, located at a distance $r$ from the centre of the clump has evolved for a time that is equal to its contraction timescale, it is collapsed into stars. Thus, the local number of cores of a given mass, at a given epoch $\tau$, is the sum of all the local populations of cores of the same mass that have formed at all epochs anterior or equal to the considered epoch with the additional step of subtracting from that sum the cores of the same mass that have readily collapsed into stars. The local populations of cores of various ages are evolved separately as they are each in a different phase of their contraction, and will collapse and form stars at various epochs. Thus, the total local number of cores of a given mass $M$, at a time $t$, is given by:

\begin{equation} 
N(r,M,t)=\sum_{\tau_{i}  \leq t} N(r,M,\tau_{i},t).
\label{eq17}
\end{equation}

We assume that only a fraction of the mass of a core ends up locked in the star. We account for this mass loss in a purely phenomenological way by assuming that the mass of a star which is formed out of a core of mass $M$ is given by M$_{\star}=\xi M$, where $\xi \le 1$. Matzner \& McKee (2000) showed that $\xi$ can vary in the range $0.25-0.75$. In this section, we describe the results for a fiducial model. In this model, the mass of the clump is taken to be M$_{clump}=10^{5}$ M$_{\odot}$, the metallicity is $Z=Z_{\odot}$, and The CFE is $CFE_{ff}=0.2$, where $t_{ff,cl}$ is the free-fall time of the clump and is given by $t_{ff,c}= (3 \pi/ 32~G \bar{\rho_{c}})^{1/2}$, and $\bar{\rho_{c}}$ the average density of the clump. The other quantities have been taken to be equal to the most commonly cited observational determinations and have been set, in the fiducial model as well as in all other  models, to the following values: $R_{c0}=0.2$ pc, $\nu=3$, $\mu=0.2$, $\kappa=0.1$. We also assume that $\xi=1/3$. Fig.~\ref{fig5} displays the time evolution of the CMF (left column) and of the IMF (right column) in the fiducial model. By $t \sim 0.3~t_{ff,c}$, the first stars form. In this model, since $\mu=0.2>~ 0$, the most massive stars form first, as the most massive cores tend to be, at any given position in the clump, more centrally peaked and thus have shorter lifetimes. As time advances, the IMF becomes fully populated. On the other hand, the CMF ceases to evolve (i.e., there is no accretion or coalescence in this model as in Dib 2007, Dib et al. 2007b; Dib et al. 2008b; and Dib et al. 2010a) as the numbers of cores that are newly formed at each position in the protocluster clump is balanced by an equal number of cores which collapses and forms stars. The final IMF of the cluster is established at $\sim 1.79~t_{ff,c}$, which corresponds to the last epoch shown in Fig.~\ref{fig5}. At this epoch, Fig.~\ref{fig6} shows that the ratio $E_{k,wind}/E_{grav}$ reaches unity (full line), and as a consequence, gas is expelled from the protocluster region. Fig.~\ref{fig7} displays the time evolution of the SFE in this model (full line). At $t=t_{exp}=1.79~t_{ff,c}$, the final value of the SFE is $SFE_{exp} \sim 9.05 \times 10^{-2}$.

In this fiducial model, we have adopted a CFE value of $CFE_{ff}=0.2$. This is an intermediate value between the values of $\sim 0.06$ and $\sim 0.33$ measured by Dib et al. (2010a) in numerical simulations of molecular clouds with two different degree of magnetisation. At first glance, it may appear that increasing the $CFE_{ff}$ by a given factor will lead to an increase in $SFE_{exp}$ by approximately the same factor. However, for a fully sampled IMF, a larger $CFE_{ff}$ value implies that a larger number of OB stars will be formed and deposit larger amounts of feedback by stellar winds in the protocluster region. This in turn leads to a faster evacuation of the gas and to a limitation of the $SFE_{exp}$. Fig.~\ref{fig6} shows that in two other models similar to the fiducial case but with a different $CFE_{ff}$, the evacuation of the gas occurs faster for the higher CFE case (i.e., case with $CFE_{ff}=0.3$) and slower for the low CFE case (i.e., $CFE_{ff}=0.1$). The time evolution of the SFE in these two additional models is compared to the fiducial case in Fig.~\ref{fig7}. The final values of the SFE at $t=t_{exp}$ are $SFE_{exp} \sim 6.59 \times 10^{-2}$, $\sim 9.05 \times 10^{-2}$, and $\sim 0.11$ for the cases with $CFE_{ff}=0.1, 0.2,$ and $0.3$ respectively. A factor of 2 and 3 increase in the $CFE_{ff}$ leads to an increase of the $SFE_{exp}$ only by factors of $\sim 1.37$ and $\sim 1.64$, respectively. This implies a strong regulation of the effect of a varying CFE by stellar feedback on the resulting $SFE_{exp}$ in protocluster clumps.      
 
 \subsection{The dependence of the star formation efficiency on metallicity}\label{metal_dep}

\begin{figure}[ht]
\includegraphics[totalheight=0.45\textheight,width=\textwidth]{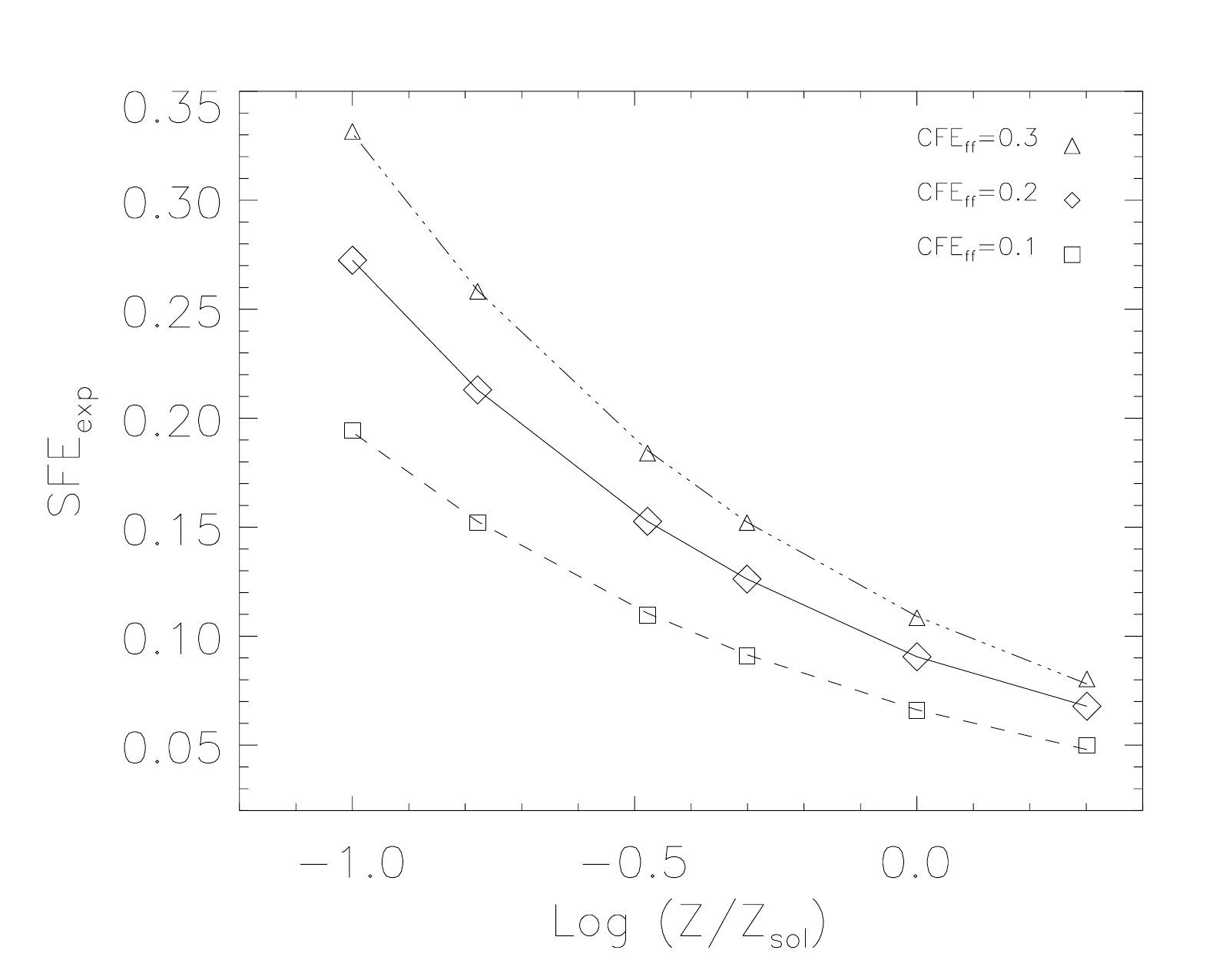} 
\caption{\label{fig9} {\small $SFE_{exp}$-Metallicity relations for clumps with the fiducial mass of $10^{5}$ M$_{\odot}$ and for different core formation efficiencies of $CFE_{ff}=0.1$, $0.2$, and $0.3$. Over-plotted to the data are fit functions (Eq.~\ref{eq25}) whose parameters are given in Dib et al. (2011). Adapted from Dib et al. (2011)}
}
\end{figure}

\begin{figure}[ht]
\includegraphics[totalheight=0.55\textheight,width=\textwidth]{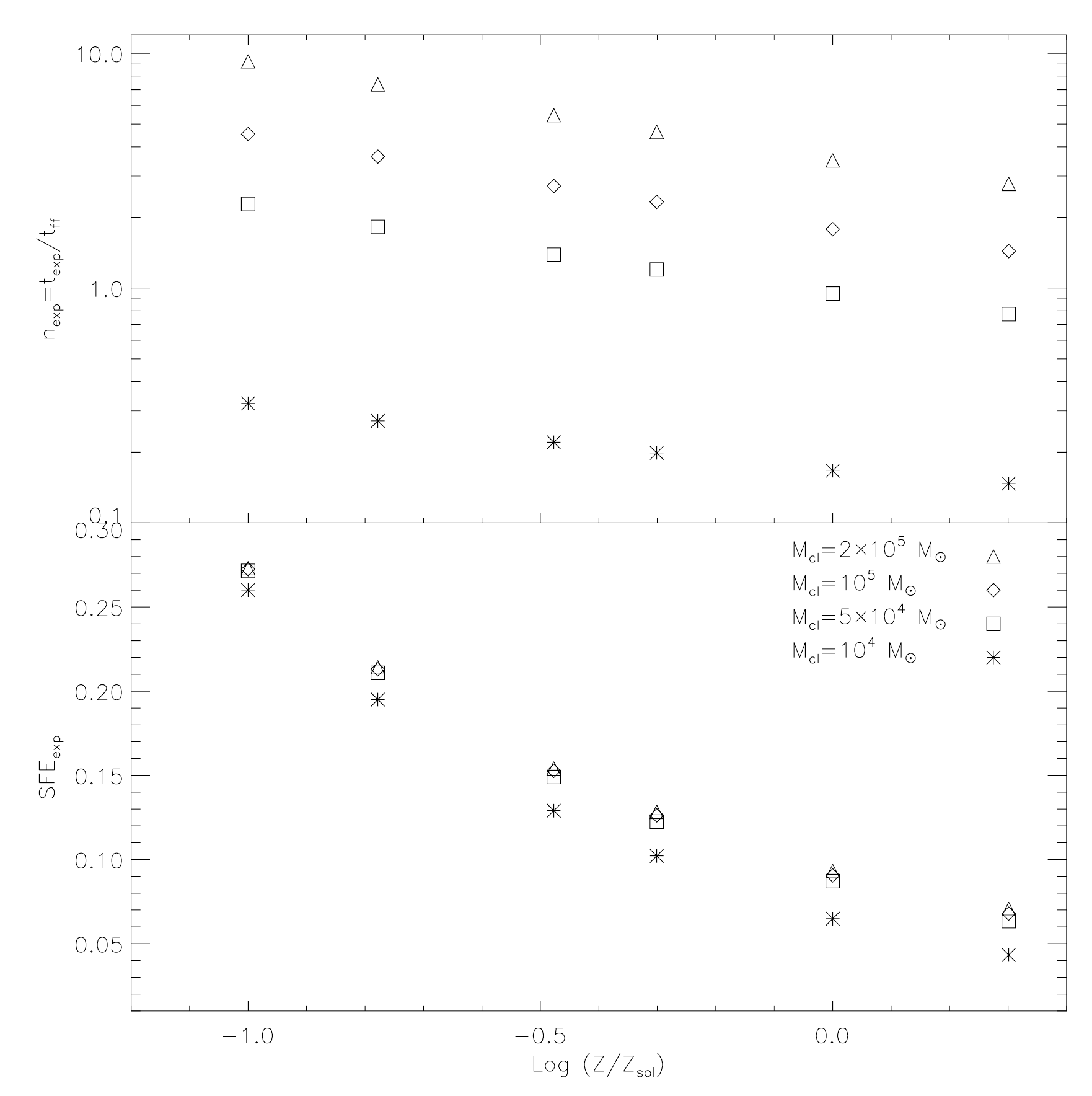} 
\caption{\label{fig10} {\small Dependence of the quantities $SFE_{exp}$ (final star formation efficiency) and $n_{exp}=t_{exp/t_{ff}}$ (ratio of the expulsion time to the free-fall time) for selected values of the protocluster clump masses and metallicies. Adapted from Dib (2011).}
}
\end{figure}

Fig.~\ref{fig8} displays the time evolution of the ratio $E_{k,wind}/E_{grav}$ in models similar to the fiducial case but with metallicities varying between $Z=0.1~Z_{\odot}$ and $Z=2~Z_{\odot}$. All six models have the fiducial value of the CFE, $CFE_{ff}=0.2$. In models with lower metallicities, the power of the stellar winds in weaker (i.e., Fig.~\ref{fig4}) and the evacuation of the gas occurs at later epochs as compared to the higher metallicity cases. In the model with $Z=2~Z_{\odot}$, the gas is expelled from the protocluster region at $t \sim 1.4~t_{ff,c}$ while in the model with $Z=0.1~Z_{\odot}$, the gas expulsion is delayed until $t \sim 4.5~t_{ff,c}$. For a given $CFE_{ff}$, longer timescales imply a larger $SFE_{exp}$. These values are plotted versus metallicity in Fig.~\ref{fig9}. A clear trend is observed in which the $SFE_{exp}$ increases with decreasing metallicity (diamonds). We fit the $SFE_{exp}-Z$ points in Fig.~\ref{fig9} with the following functional form:

 \begin{equation} 
SFE_{f}= C_{Z}~e^{-\frac{1}{\tau_{Z}} \log \left(\frac{Z}{Z_{\odot}} \right)}.
\label{eq18}
\end{equation}
   
\noindent Fo the case case with $CFE_{ff}$=0.2, the fit parameters are $C_{Z}=0.091 \pm 6.8\times 10^{-4}$ and $\tau_{Z}=0.91\pm 7.9 \times 10^{-3}$. We have also repeated the calculations at the various metallicities using the additional values of $CFE_{ff}=0.1$ and $CFE_{ff}=0.3$. The dependence of the $SFE_{exp}$ on metallicity in these cases are also displayed in Fig.~\ref{fig9}. Fig.~\ref{fig9} shows that the $SFE_{exp}$ varies typically by a factor of $\sim 1.3-1.4$ and $\sim 1.6-1.7$ for variations of the CFE by factors of 2 and 3, respectively. This argues for a strong regulation of the SFE by feedback. 

\section{The Kennicutt-Schmidt laws in the feedback regulated model of star formation}

Using the above described model, it is possible to derive a metallicity dependent star formation rate (SFR) at a given surface density $\Sigma_{g}$. For a summary on the recent observational determinations of the star formation laws see Kennicutt (2008), Elmegreen (2011), and Dib (2011). The star formation rate surface density in the feedback regulated mode of star formation is given  by:

\begin{equation} 
\Sigma_{SFR}=\Sigma_{g}~f_{H2} \frac{\left<SFE_{exp}\right>} {\left<t_{exp}\right>},
\label{eq19}
\end{equation}  

\noindent where $\left<SFE_{exp}\right>$ and $\left<t_{exp}\right>$ are, respectively, the characteristic $SFE_{exp}$ and the epoch at which gas is expelled from the protocluster region for the clump mass distribution associated with a given $\Sigma_{g}$. Writing $\left<t_{exp}\right>$ in terms of the clumps free-fall time $\left<t_{ff}\right>$, Eq.~\ref{eq2} becomes:

\begin{eqnarray}
\Sigma_{SFR}=\Sigma_{g}~f_{H_{2}} \frac{\left<SFE_{exp}\right>} {\left<n_{exp}\right>} \frac{1}{\left<t_{ff}\right>} \nonumber \\
                                 ~=\Sigma_{g}~f_{H_{2}} \frac{\left<f_{\star,ff}\right>}{\left<t_{ff}\right>}.
\label{eq20}                  
\end{eqnarray}

\noindent where $f_{\star,ff}$ is the dimensionless star formation efficiency and which corresponds to the mass fraction of the molecular gas that is converted into stars per free-fall time $t_{ff,c}$ of the clumps and. $\left< f_{\star,ff}\right>$ and $\left< t_{ff}\right>$ represent characteristic values of $f_{\star,ff}$ and $t_{ff,c}$ for the spectrum of clump masses found in the giant molecular for a given value of $\Sigma_{g}$. The quantity $f_{H_{2}}$ is the mass fraction of the total gas that is in molecular form. In this work, we use the functional form of $f_{H_{2}}$ obtained by Krumholz et al. (2009a) who derived $f_{H_{2}}$ as a function of the gas surface density and metallicity (see their paper or Dib 2011 for the detailed formula). $\left<t_{ff,c}\right>$ can be approximated by the free-fall time of the clump with the characteristic mass $t_{ff,c} (M_{char})=8 \Sigma_{cl}^{' -3/4} M_{char,6}^{1/4}$ Myr where $M_{char,6}=M_{char}/10^{6}$ M$_{\odot}$. The characteristic mass $M_{char}$ is given by :

\begin{figure}[ht]
\includegraphics[totalheight=0.4\textheight,width=\textwidth]{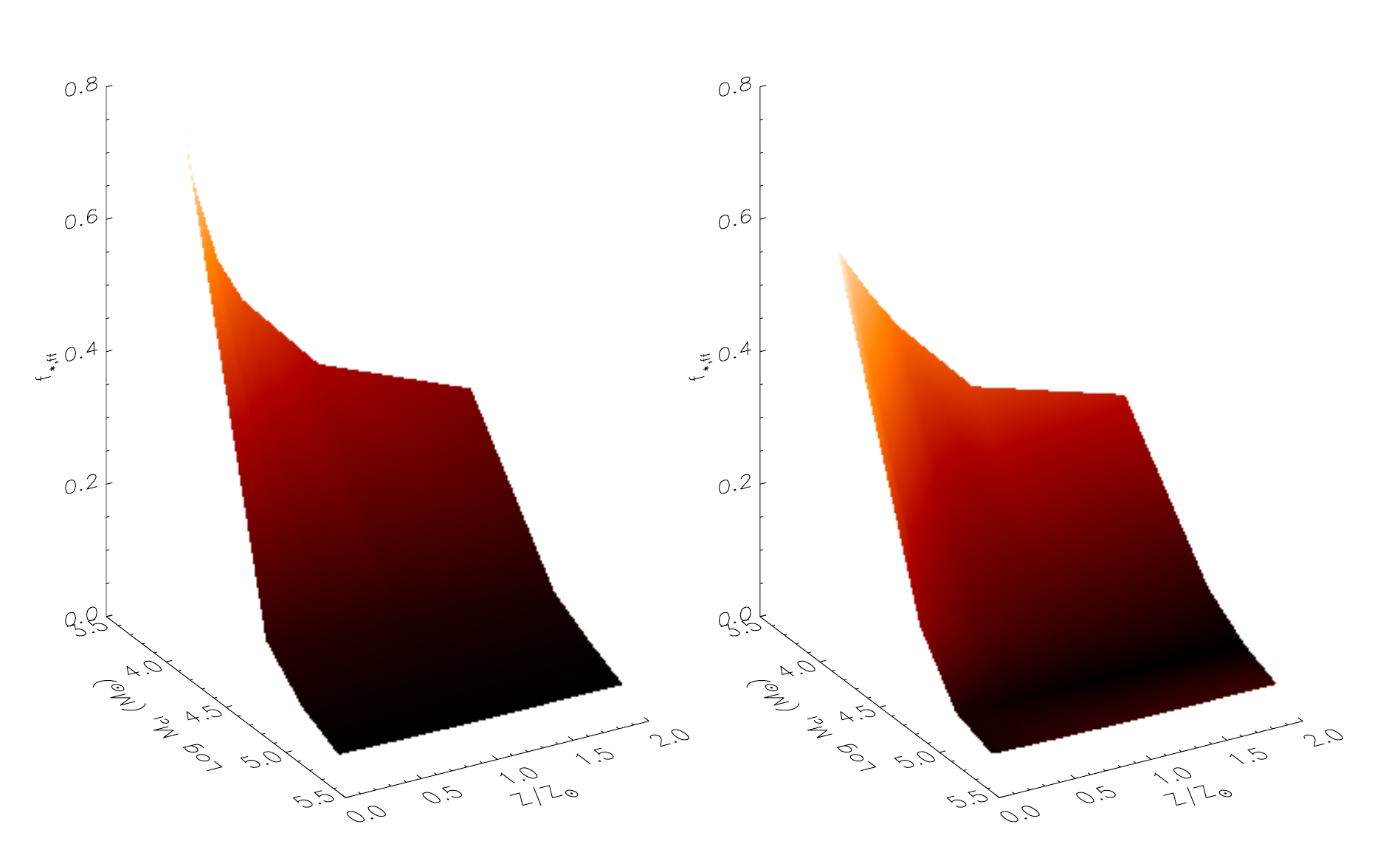} 
\caption{\label{fig11} {\small Star formation efficiency per unit free-fall time in the protocluster clump in the metallicity-dependent feedback model. The star formation efficiencies per free-fell use a core-to-star efficiency conversion factor of 1/3. The left panel displays $f_{\star,ff}$ as a function of both $M_{cl}$ and $Z^{'}=Z/Z_{\odot}$ in the original data, while the right panel displays the analytical fit function to this data set given in Eq.~\ref{eq8}. Adapted from Dib (2011).}
}
\end{figure}

\begin{equation}
 M_{char}= \int_{M_{cl,min}}^{max(M_{cl,max},M_{GMC})} M_{cl} N(M_{cl}) dM_{cl}, 
 \label{eq21}
\end{equation}

\begin{figure}[ht]
\includegraphics[totalheight=0.45\textheight,width=\textwidth]{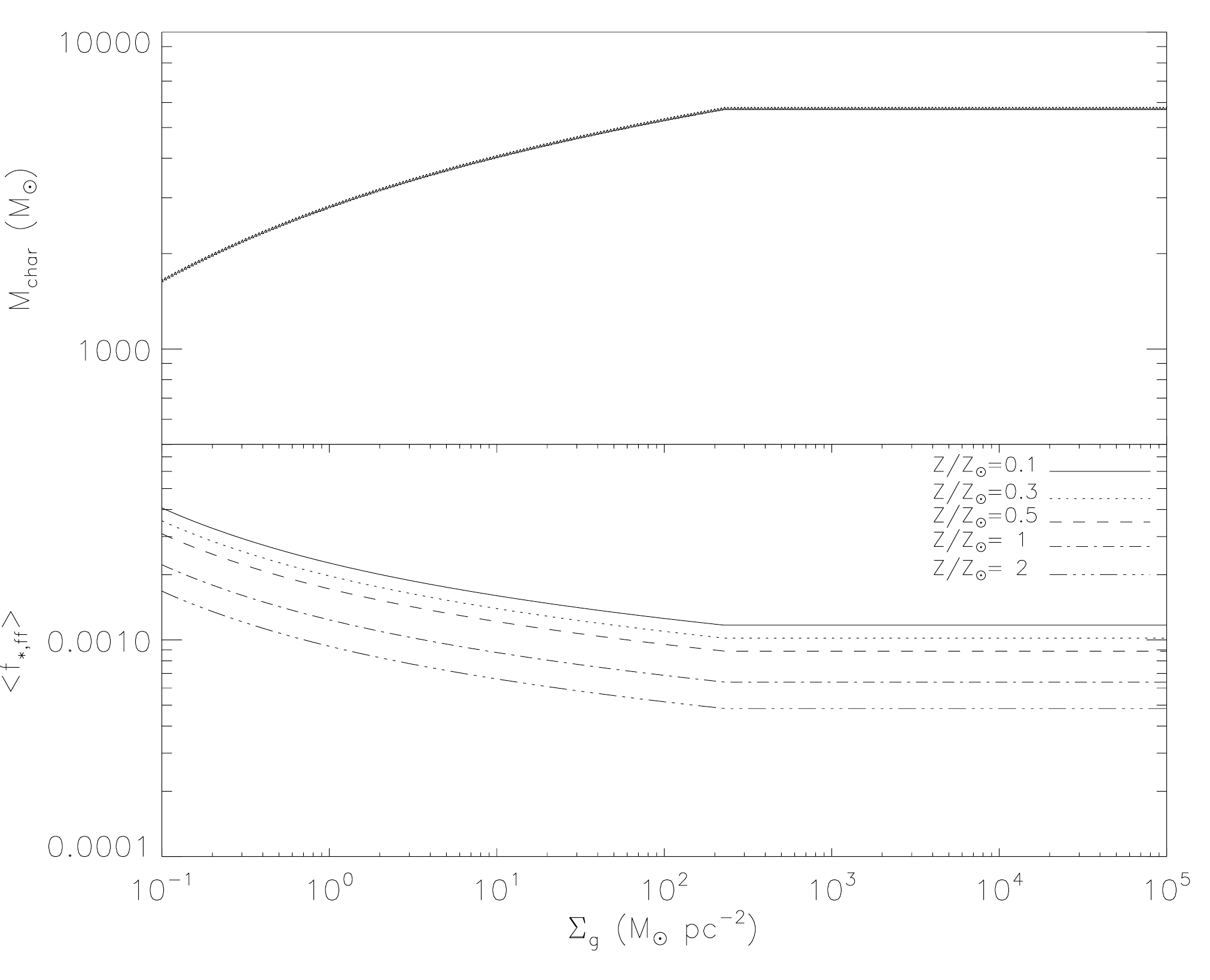} 
\caption{\label{fig12} {\small Characteristic clump mass as a function of the gas surface density (Eq.~\ref{eq4}, top panel) and the star formation efficiency per unit free-fall time in this feedback regulated model of star formation. Adapted from Dib (2011).}
}
\end{figure}

\noindent where $N (M_{cl})$ is the mass function of protocluster forming clumps which we take to be $N (M_{cl})=A_{cl} M_{cl}^{-2}$, and $A_{cl}$ is a normalisation coefficient given by $A_{cl} \int_{M_{cl,min}}^{max(M_{cl,max},M_{GMC})} N(M_{cl}) dM_{cl}= \epsilon$, where $0 < \epsilon < 1$ is the mass fraction of the GMCs that is in protocluster clumps at any given time. In this work we use $\epsilon=0.5$. The minimum clump mass $M_{cl,min}$ is taken to be $2.5 \times 10^{3}$ M$_{\odot}$ (this guarantees, for final SFEs  in the range of 0.05-0.3 a minimum mass for the stellar cluster of $\sim 50$ M$_{\odot}$) and the maximum clump mass is $10^{8}$ M$_{\odot}$. The characteristic GMC mass is determined by the local Jeans mass and is given by:

\begin{equation} 
M_{GMC}=37 \times 10^{6} \left(\frac{\Sigma_{g}}  {85~\rm{M_{\odot}~pc^{-2}}} \right) {\rm M_{\odot}}. 
\label{eq22}
\end{equation} 

Fig.~\ref{fig12} (top) displays $M_{char}$ as a function of $\Sigma_{g}$. Fig.~\ref{fig10} displays the dependence of $SFE_{exp}$ and $t_{exp}$ on clump mass and metallicity for a series of models in which the $CFE_{ff}=0.2$. The $SFE_{exp}$ depends strongly on metallicity but weakly on mass whereas $t_{exp}$ displays a clear dependence on both quantities. The quantity $f_{\star,ff}=SFE_{exp}/n_{exp}$ is displayed in Fig.~\ref{eq11} (left panel) as a function of mass and metallicity ($Z^{'}=Z/Z_{\odot}$). A fit to the $f_{\star,ff}$ ($M_{cl},Z^{'}$) data points with a 2-variables second order polynomial yields the following relation (shown in Fig.~\ref{fig11}, right panel):

\begin{eqnarray}
f_{\star,ff}(M_{cl},Z^{'})=  11.31-4.31{\rm log(M_{cl})} + 0.41 {\rm [log(M_{cl})]^{2}}  \nonumber \\ 
           - 8.28 Z^{'} + 3.20 Z^{'} {\rm log (M_{cl})} - 0.32 Z^{'} {\rm [log(M_{cl})]^{2}}  \nonumber \\
          + 2.30 Z^{'2} - 0.89 Z^{'2} {\rm log(M_{cl})} + 0.08 Z^{'2} {\rm [log(M_{cl})]^{2}}.
\label{eq23}
\end{eqnarray} 

\begin{figure}[ht]
\includegraphics[totalheight=0.5\textheight,width=\textwidth]{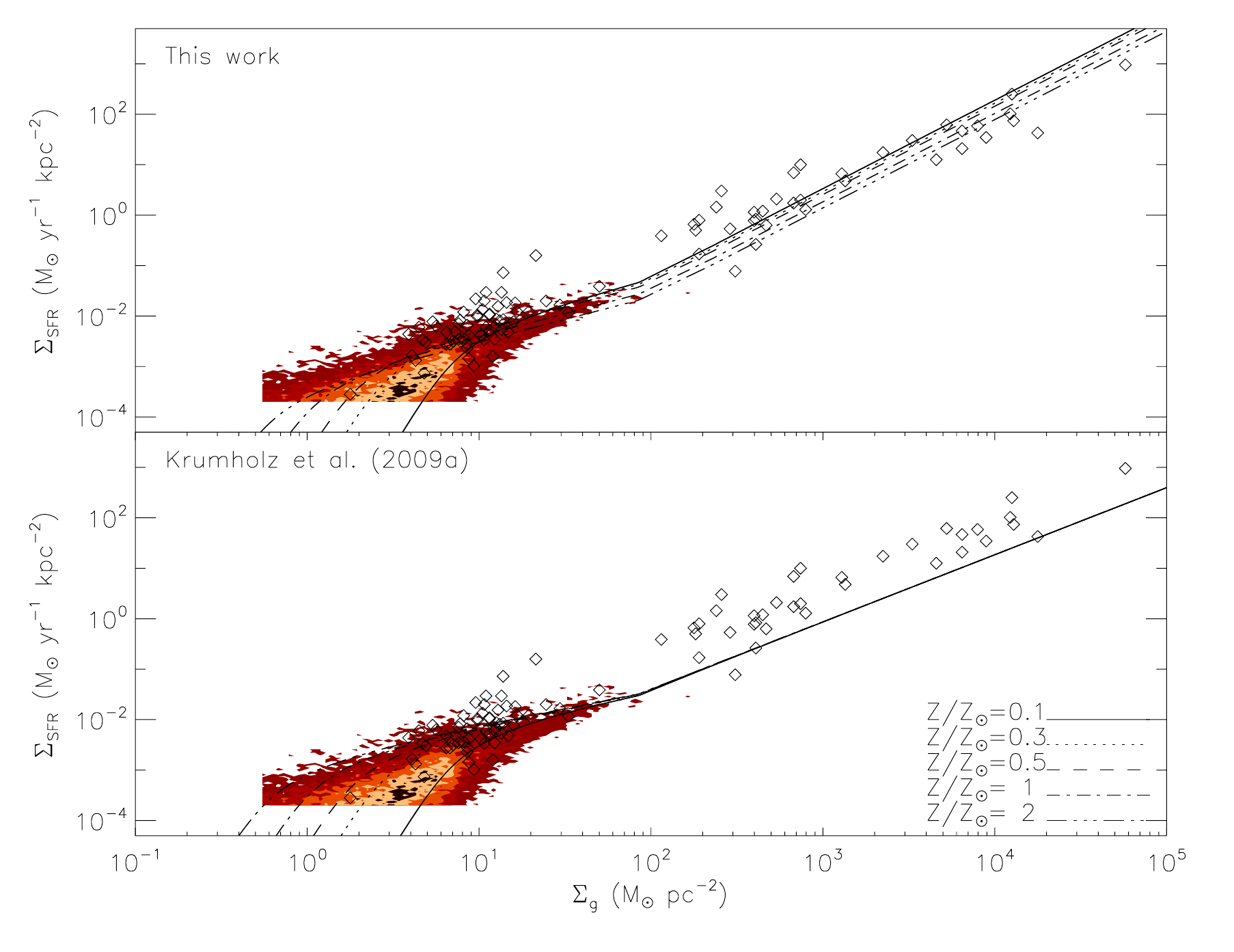} 
\caption{\label{fig13} {\small Star formation laws in the feedback-regulated star formation model (this work, top panel), and in the Krumholz et al. (2009a) model (bottom panel). Overplotted to the models are the normal and starburst galaxies data of Kennicutt (1998) and the combined sub-kpc data (4478 subregions in total) for 11 nearby galaxies from Bigiel et al. (2008,2010). The Bigiel et al. (2008,2010) data is shown in the form of a 2D histogram with the color coding corresponding, from the lighter to the darker colors to the 1,5,10,20, and 30 contour levels. The displayed theoretical models cover the metallicity range $Z^{'}=Z/Z_{\odot}=[0.1,2]$. Adapted from Dib (2011).}
}
\end{figure}

Using Eq.~\ref{eq23}, it is then possible to calculate $\left<f_{\star,ff}\right>$:  

\begin{equation}
\left<f_{\star,ff}\right> (Z^{'}, \Sigma_{g}) = \int_{M_{cl,min}}^{max(M_{cl,max},M_{GMC})} f_{\star,ff} (M_{cl},Z^{'}) N(M_{cl}) dM_{cl}. 
\label{eq24}
\end{equation}

Fig.~\ref{fig12} (bottom) displays $\left<f_{\star,ff}\right>$ ($Z^{'},\Sigma_{g}$) for values of $Z^{'}$ in the range [$0.1-2$]. We assume that there is a critical value of $\Sigma_{g}= 85$ M$_{\odot}$ pc$^{-2}$ below which clumps are pressurised by their internal stellar feedback, such that $\Sigma_{cl}=\Sigma_{g,crit}$ where $\Sigma_{g} < \Sigma_{g,crit}$ and $\Sigma_{cl}=\Sigma_{GMC}=\Sigma_{g}$ when $\Sigma_{g} \geq \Sigma_{g,crit}$. With the above elements, the star formation law can be re-written as:

\begin{eqnarray}
\Sigma_{SFR} &=& \frac{8} {10^{6}} f_{H_{2}} (\Sigma_{g},c,Z^{'}) \Sigma_{g}  \nonumber \\
  \times & & \left \{\begin{array} {cc}  \frac{\left<f_{\star,ff}\right> (Z^{'})} {M_{char,6}^{1/4}} &  ; \frac{\Sigma_{g}} {85~{\rm M_{\odot}~pc^{-2}}} < 1\\
  \frac{\left<f_{\star,ff}\right> (Z^{'})} {M_{char,6}^{1/4}} \left(\frac{\Sigma_{g}}{85~{\rm M_{\odot} pc^{-2}}}\right)^{3/4}  & ; \frac{\Sigma_{g}}{85~{\rm M_{\odot}~pc^{-2}}} \geq 1 \end{array} \right \},
\label{eq25}
\end{eqnarray}

\noindent where $\Sigma_{SFR}$ is in M$_{\odot}$ yr$^{-1}$ kpc$^{-2}$, $M_{char}$ is given by Eq.~\ref{eq21}, and $\left<f_{\star,ff}\right>$ by Eqs.~\ref{eq23} and \ref{eq24}. Fig.~\ref{fig13} (top panel) displays the results obtained using Eq.~\ref{eq25} for $\Sigma_{g}$ values starting from low gas surface densities up to the starburst regime. The results are calculated for the metallicity values of $Z^{'}=[0.1,0.3,0.5,1,2]$. The results are compared to the sub-kpc data of Bigiel et al. (2008,2010) and to the normal and starburst galaxies results of Kennicutt (1998) (Fig.~\ref{fig13}, bottom panel displays the results of Krumholz et al 2009 which is based on the regulation of star formation by turbulence in GMCs). Our models fits remarkably well the observational results over the entire range of surface densities. Furthermore, the segregation by metallicity extends beyond the low surface density regime up to the starburst regime where a segregation in metallicity of $\sim 0.5$ dex is observed, in contrast to the Krumholz et al. (2009b) models which do not contain a metallicity dependence in the intermediate to high surface density regimes. Furthermore the solar metallicity curve in our model overlaps with a significant fraction of the sub-regions in the data of Bigiel et al. (2008,2010) in contrast to the Krumholz et al. model.        
 
\section{CONCLUSIONS}\label{conc}

We have described a model for star formation in protocluster clumps of difference metallicities. The model describes the co-evolution of the dense core mass function and of the IMF in the clumps. Cores form uniformly over time in the clumps following a prescribed core formation efficiency per unit time. Cores contract over timescales which are a few times their free fall time before they collapse to form stars. Feedback from the newly formed OB stars ($> 5$ M$_{\odot}$) is taken into account and when the ratio of the cumulated effective kinetic energy of the winds to the gravitational energy of the system (left over gas+stars) reaches unity, gas is expelled from the clump and further core and star formation are quenched. The radiation driven winds of OB stars are metallicity dependent. Metal rich OB stars inject larger amount of energy into the clump than their low metallictiy counterparts and thus help expel the gas on shorter timescales. This results in reduced final star formation efficiencies in metal rich clumps in comparison to their low metallicity counterparts. Both the final star formation efficiency and the gas expulsion timescales are combined for a grid of clump models with different masses and metallicities in order to calculate the star formation efficiency per unit time ($f_{\star,ff}$) in this feedback regulated model of star formation. We calculate the characteristic value of $f_{\star,ff}$ for a clump mass distribution associated with a gas surface density, $\Sigma_{g}$. This is combined with a description of the molecular mass fraction as a function of $\Sigma_{g}$ and the assumption that there is a critical surface gas density ($\Sigma_{g}=85$ M$_{\odot}$ pc$^{-2}$) above which the protocluster clumps and their parent giant molecular clouds switch from being pressurised from within by stellar feedback to being confined by the external interstellar medium pressure. The combination of these three elements allows us to construct the star formation laws in galaxies going from low gas surface densities up to the starburst regime. Our models exhibit a dependence on metallicity over the entire range of considered gas surface densities and fits remarkably well the observational data of Bigiel et al. (2008,2010) and Kennicutt (1998). This dependence on metallicity of the KS relation may well explain the scatter (or part of it) that is seen in the observationally derived relations.  

%
%
\small  
%
\section*{Acknowledgments}  
I would like to thank the members of the organising committees for setting up a very inspiring conference and Subhanjoy Mohanty, Jonathan Braine, George Helou, Laurent Piau for useful discussions and collaboration on this work. Support for this work has been provided by STFC grant ST/H00307X/1.
%
%

%

%
\end{document}